\documentclass[12pt]{article}
\title{
Magnetic moment of $\Delta^{++}$ baryon in QCD string approach\\
 }
\author{B.O. Kerbikov\\ State Research
Center\\Institute of Theoretical and Experimental Physics, \\ Moscow,
Russia}
 \date{}
  \newcommand{\be}{\begin{equation}}
\newcommand{\ee}{\end{equation}}
 
\def\fun#1#2{\lower3.6pt\vbox{\baselineskip0pt\lineskip.9pt
\ialign{$\mathsurround=0pt#1\hfil
##\hfil$\crcr#2\crcr\sim\crcr}}}
\newcommand{\Del}{\Delta^{++}}

\newcommand{\ver}{\mbox{\boldmath${\rm r}$}}

\newcommand{\ves}{\mbox{\boldmath${\rm s}$}}

\newcommand{\vexi}{\mbox{\boldmath${\rm \xi}$}}
\newcommand{\veta}{\mbox{\boldmath${\rm \eta}$}}

\newcommand{\lan}{\langle}
\newcommand{\ran}{\rangle}

\begin{document}
\maketitle

\begin{abstract}
Magnetic moments (m.m.) of the  $\Delta^{++}$ baryons is computed
within the new approach based on the QCD string Hamiltonian.
   The string tension $\sigma$ is only dimensionful quantity forming
m.m. of both nucleon and $\Del$, but color Coulomb and spin-spin
interactions cancel each other in nucleon m.m. while in $\Del$ they
add coherently. The result $\mu_{\Del} = 4.36\mu_N$ is in good
agreement with experimental data.

   \end{abstract}

 Recently in  \cite{1} a new approach to
evaluation of the baryon m.m. has been proposed. It is based on the
Feynman-Schwinger (world-line) representation of the 3q Green's
function and yields remarkably simple expressions for the m.m.
through the only fundamental parameter --string tension $\sigma$ (and
strange quark current mass for strange baryons.)
  In \cite{1} calculations have been performed for the octet baryons
  and for $\Omega^-$. The present letter aims at the calculation of
  the $\Del$ m.m. and some amendments to our previous
  treatment of $\Omega^-$.

  The case of $\Del$ is a particular one both from experimental and
  theoretical sides. Experimentally m.m. of $\Del$ was measured
  rather recently and is still subject to substantial uncertainties
  \cite{2}. On the theoretical side the calculation of the  $\Del$
  m.m. causes serious  difficulties in various approaches and calls
  for introduction of  several additional parameters (see e.g. recent
  paper \cite{3} for a brief  review of the current status of the sum
  rules approach to the problem) As will be seen in what follows the
  treatment of the $\Del$ in our approach requires taking into
  account hyperfine and color Coulomb interactions. This was not the
  case for octet baryons considered in \cite{1} since for them
  Coulomb and spin-spin terms extinguished  each other \cite{1}.
   For $\Omega^-$ baryon also considered in \cite{1} the situation is
   the following. As for other baryons the m.m. of $\Omega^-$ is
   predominantly controlled by string tension $\sigma$, next comes
   the contribution from the strange quark current mass $m_s$, and
   then Coulomb and hyperfine terms contribute. In \cite{1} the first
   two factors ($\sigma$ and $m_s$) were considered while the present
   work deals in addition with the two remaining terms  listed above.
   This does not alter the value of the $\Omega^-$ m.m. obtained in
   \cite{1} but allows to take somewhat smaller value of $m_s$ than
   that used in \cite{1}.

   We start with a recapitulation of the very  few key points of the
   approach to baryon m.m. developed in \cite{1}. For the $3q$
   Green's function one can write the following Feynman--Schwinger
   (world-line representation \cite{4}-\cite{6})
\be
G^{(3q)}(X,Y)=\int\prod^3_{i=1} ds_i Dz^{(i)}_\mu e^{-K}\lan W_3
(X,Y)\ran,
\label{1}
\ee
where $X;Y=x^{(i)}; y^{(i)}, i=1,2,3$, and the integration is along
the path $z^{(i)}_\mu(s_i)$ of the $i$-th quark with $s_i$ playing
the role of the proper "time" parameter along the path, and
\be
K=\sum^3_{i=1} \left (m_i^2s_i+\frac14 \int^{s_i}_0\left
(\frac{dz^{(i)}_\mu}{d\tau_i}\right ) d\tau_i\right).
\label{2}
\ee
Here $m_i$ is the current quark mass and the three-lobed Wilson loop
is a product of the three parallel transporters \cite{4}-\cite{6}.
The standard approximation of the QCD string approach is the minimal
area law for $\lan W_3\ran$
 \be
  \lan W_3\ran =\exp (-\sigma
\sum^3_{i=1}S_i), \label{3} \ee where $S_i$  is the minimal area of
one loop. The next step is to calculate the quark  constituent mass
$\nu_i$ in terms of its current mass $m_i$ and string tension
$\sigma$. To this end one connects the proper and real times via \be
ds_i=\frac{dt}{2\nu_i(t)},
\label{4}
\ee
where $t=z^{(i)}_4(s_i)$ is a common c.m. time on the hypersurface
$t=const $ \cite{4,5}. The new entity, $\nu_i(t)$, being determined
from the condition of the minimum of the corresponding Hamiltonian
(see below) plays the role of the quark constituent mass (see
\cite{1,4} for details). Referring to \cite{1,4} for the derivation
of the Hamiltonian from the above defined Green's function $G^{(3q)}$
we write down the final expression containing $\nu_i$ as parameters
\be
H=\sum^3_{k=1}\left(\frac{m^2_k}{2\nu_k}+\frac{\nu_k}{2}\right)+\frac{1}{2m}\left(
-\frac{\partial^2}{\partial\xi^2}-\frac{\partial^2}{\partial\eta^2}\right)+ \sigma
\sum^3_{k=1}|\ver^{(k)}|,
\label{5}
\ee
where $\vexi$ and $\veta$ are three-body Jacoby coordinates defined
as in \cite{6,7}, $m$ is  an arbitrary mass parameter which ensures
correct dimensions and drops out of final expressions, and
$|\ver^{(k)}|$ is the distance from the $k$-sh  quark to the
string-junction positions which was for simplicity taken coinciding
with the c.m. point. It is at this point worth stressing that the QCD
string model outlined above is a fully relativistic string model for
light current masses, and the "nonrelativistic" appearance of the
Hamiltonian (\ref{5}) is a consequence of the rigorous einbein
formalism \cite{8}.

Tooled with the Hamiltonian (\ref{5}) one can use the standard
hyperspherical formalism \cite{7}, introduce hyperradius
$\rho^2=\xi^2+\eta^2$, and write the  following eigenwalue equation
(quarks with equal masses are considered)
\be
\frac{d^2\chi}{d\rho^2} +2\nu\{E_n-W(\rho)\}\chi(\rho)=0,
\label{6}
\ee
\be
W(\rho)=b\rho+\frac{d}{2\nu\rho^2},
b=\sigma\sqrt{\frac23}\frac{32}{5\pi}, d=15/4.
\label{7}
\ee
The baryon mass $M_n(\nu)$ is equal to
\be
M_{n}(\nu)=\frac{3m^2}{2\nu} +\frac32\nu +E_n(\nu).
\ee
According to the QCD string prescription \cite{4}-\cite{6} the value
of $\nu$ is  determined as a stationary point of $M_n(\nu)$:  \be
\frac{\partial M_n}{\partial\nu}=0.
\label{9}
\ee
In passing from (\ref{4}) to (\ref{9}) we have changed from $\nu(t)$
depending on the trajectory to the operator $\nu$ and finally to the
constant $\nu$ to be found from the  minimum condition (\ref{9}) --
see \cite{4}-\cite{6} for details.

The perturbative gluon exchanges and spin--dependent terms can be
selfconsistently included into the above picture.
Including the Coulomb term and passing to dimensionless quantities
$x,\varepsilon_n$ and $\lambda$ defined as
\be
x=(2\nu b)^{1/3}\rho, ~~\varepsilon_n=\frac{2\nu E_n}{(2\nu
b)^{2/3}},~~ \lambda=\alpha_s=\frac83\left
(\frac{10\sqrt{3}\nu^2}{\pi^2\sigma}\right)^{1/3}\equiv \tilde
\lambda\left( \frac{\nu^2}{\sigma}\right)^{1/3},
\label{10}
\ee
where $\alpha_s$ is the strong coupling constant, one arrives at the
following reduced equation
\be
\left\{-\frac{d^2}{dx^2} +x+\frac{d}{x^2}
-\frac{\lambda}{x}-\varepsilon_n(\lambda)\right\}=\chi(x)=0.
\label{11}
\ee
Then (\ref{9}) yields the equation defining the quark  dynamical mass
$\nu$
\be
\varepsilon_n(\lambda)\left( \frac{\sigma}{\nu^2}\right)^{2/3}
\left \{
1+\frac{2\lambda}{\varepsilon_n(\lambda)}
\left\vert\frac{d\varepsilon_n}{d\lambda}
\right\vert\right\}+\frac{9}{16}\left( \frac{75\pi^2}{2}\right)^{1/3}
\left(\frac{m^2}{\nu^2}-1\right)=0.
\label{12}
\ee

At this point the essential difference of $\Del$ from $p$ or $n$
arises. It concerns the interplay of the Coulomb and spin-spin
interaction (the later is not yet included into (\ref{11}) and
(\ref{12})). The spin-spin interaction in baryon made of equal mass
quarks results in the shift of $E_n$ equal to
\be
\delta E_n=\frac{16}{9} \frac{\alpha_s}{\nu^2} \sum_{i>j}
\ves_i\ves_j\delta(\ver_{ij}).
\label{13}
\ee
For $\Del$ summation over ($i,j)$ yields a factor 3/4 corresponding
to positive interaction energy \cite{9}, i.e. in $\Del$  spin-spin
interaction acts coherently with the Coulomb one. In nucleon the
corresponding factor is  --4/3 resulting in reverse situation.  This
distintion realizes in the fact that the masses of quarks forming the
nucleon remain "unrenormalized" due to Coulomb and spin-spin
interaction \cite{1} while the masses of quarks forming $\Del$
substantially increase as shown below.

To include the term (\ref{13}) into equation (\ref{12}) one has to
smear the delta functions over small regions. This procedure has been
done by two independent methods in \cite{10} with the result
$\lan\delta(\ver_{ij})\ran=\delta_0\nu^{3/2}, \delta_0=2.64\cdot
10^{-2}$ GeV$^{3/2}$. With spin-spin interaction included Eq.
(\ref{12}) for $\Del$ takes the form
\be
\nu^2-\frac{16}{9} \left( \frac{2}{75 \pi^2}\right)^{1/3}
\varepsilon(0) \sigma^{2/3} \nu^{2/3}
\left[1+\frac{\tilde\lambda}{\varepsilon
(0)}\left\vert\frac{d\varepsilon}{d\lambda}\right\vert_{\lambda=0}\left(
\frac{\nu^2}{\sigma}\right)^{1/3}\right]-\frac{4\pi}{9} \alpha_s
\delta_0\nu^{1/3}=0.
\label{14}
\ee
Only terms linear in Coulomb coupling constant $\tilde \lambda$ are
kept in (\ref{14}), the corresponding expansion parameter is
$\tilde\lambda /\varepsilon(0)\simeq 1/4$. Eq(\ref{14}) has been
solved for the following set of parameters
\be
\sigma=0.15 {\rm GeV}^2,~~ \alpha_s=0.39.
\label{15}
\ee
          The string tension value (\ref{15}) which is smaller than
in meson case is in line with baryon calculations by Capstick and
Isgur \cite{11}. A similar smaller value of $\sigma$ is implied by
recent lattice calculations by Bali \cite{12}. Solving (\ref{14}) with
the above set of parameters one gets
\be
\nu_{\Delta}=0.43 {\rm GeV},
\label{16}
\ee
which should be compared to $\nu_N=c\sqrt{\sigma}=0.37$ GeV, where
$C=0.957$ is a constant calculated in \cite{1}.

Now we turn directly to $\Del$ magnetic moment. The magnetic momet
interaction term is included into $G^{(3q)}$ in a  straightforward
way \cite{1,4}-\cite{6} resulting in
\be
\mu_{\Del}=3\lan
\psi_{\Del}\left\vert\frac{e_3\sigma_z^{(3)}}{2\nu_{\Del}}\right\vert\psi_{\Del}\ran,
\label{17}
\ee
where  $e_3=e_u=2/3$.
  The structure of this matrix element with $\nu_{\Del}$ in the
  denomitor may be considered as an additional evidence that the
  quantity $\nu$ first introduced by (\ref{4}) has the physical
  meaning of the quark constituent mass.

  The calculation of the matrix element (\ref{17}) is trivial
  provident one considers only totally symmetric coordinate wave
  functions, i.e. the lowest hyperspherical harmonic. It is known
  that the contribution of higher harmonics into normalization does
  not exceed few percent \cite{7,13}. Then (\ref{17}) yields
  \be
  \mu_{\Del}=\frac{2m_p}{\nu_{\Delta} }\simeq 4.36\mu_N,
  \label{18}
  \ee
  which is in good agreement with the experimental value $\mu_{\Del}=
  (4.52\pm 0.50\pm 0.45)\mu_N$ \cite{2}.

  The treatment of the $\Del$ baryon presented above makes  it
  possible to reexamine the case of $\Omega^-$  baryon considered in
  \cite{1}  and  to improve upon the value of the strange quark
  current mass used in \cite{1}. Namely in \cite{1} the constituent
  mass  $\nu_\Omega$ was calculated using Eq.(\ref{12})   above and
  then m.m. of $\Omega^-$ was obtained as a matrix element similar to
  (\ref{17}). Coulomb and spin-spin terms in  $\Omega^-$ were
  neglected since their contribution is smaller than that of the
  strange quark constituent mass $m_s$. If following the lines
  outlined above these hitherto omitted terms are included into the
  treatment of the $\Omega^-$ m.m. all the values of the   baryon
  m.m. presented in \cite{1} (see also Table 1 below)  remain
  unchanged but they are reproduced with smaller values of $m_s$,
  namely with $m_s=0.15$ GeV instead of $m'_s=0.245$  GeV used in
  \cite{1}. The new value of $m_s$ agrees with the result
  $m_s=0.175\pm 0.025$ GeV deduced by  Leutwyler \cite{14}.

  In Table 1  we summarize the results on baryon m.m. obtained in the
  present work and in \cite{1}. The typical deviations from the
  experimental values are about 10\% which is remarkably successful
  keeping in mind plentiful possible corrections (meson exchanges,
  higher harmonics, etc).

  The author is especially indebted to Yu.A.Simonov for formulating
  the problem, discussions and remarks.
  The author is grateful to Yu.S.Kalashnikova for numerous enlighting
  discussions and suggestions and to A.M.Badalian  and N.O.Agasian
  for useful remarks.

  The financial support of the grant RFFI 00-02-17836 and RFFI
  96-15-96740 are gratefully acknowledged.

        \newpage

{\bf Table 1.} Magnetic moments of baryons (in nuclear magnetons)
   in
  comparison with experimental data from PDG \cite{}
   \begin{center}

\begin{tabular}{|l|l|l|l|l|l|l|l|l|l|l|}\hline
Baryon&$p$&$n$&$\Del$&
$\Lambda$&$\Sigma^-$&$\Sigma^0$&$\Sigma^+$&
$\Xi^-$&$\Xi^0$&$\Omega^-$\\\hline
Ref.\cite{1} and &&&&&&&&&&\\
present work&2.54 &-1.69 &4.36&-0.69& -0.90 &0.80
&2.48&-0.63&-1.49&-2.04\\\hline
Experiment&2.79 &-1.91&4.52&-0.61&-1.16&&2.46&-0.65&-1.25&-2.02\\
\hline \end{tabular} \end{center}


\begin{thebibliography}{99}

\bibitem{1}
B.O.Kerbikov and Yu.A.Simonov, Baryon magnetic moments in
the QCD string approach, hep-ph/0001243.

\bibitem{2}
PDG group, Review of particle Physics, The European Physical
Journal {\bf C3} (1998)

\bibitem{3}
Y.M.Aliev, A.\"{O}zpineci, M.Savci, Magnetic moments of
$\Delta $ baryons in light cone QCD sum rules,
hep-ph/0002228.


\bibitem{4}
 Yu.A.Simonov, Nucl. Phys. {\bf B307} (1988) 512; Phys. Lett. {\bf
 B226} (1989) 151; hep-ph/9911237.

   \bibitem{5}
    A.Yu.Dubin, A.B.Kaidalov and Yu.A.Simonov, Phys. Lett. {\bf B323
     } (1994) 41


   \bibitem{6}
   M.Fabre de la Ripelle and Yu.A.Simonov, Ann. Phys. (N.Y.) {\bf 212}
  (1991) 235



  \bibitem{7}
   Yu.A.Simonov, Yad. Fiz. {\bf 3} (1966) 630; \\
   A.M.Badalian, Yu.A.Simonov, Yad.  Fiz.
    {\bf 3} (1966) 1032;\\
    F.Calogero, Yu. A.Simonov Phys. Rev. {\bf 183} (1968) 869.


    \bibitem{8}
    L.Brink, P.Di Vecchia, P.Howe, Nucl. Phys. {\bf B118} (1977)
    76;\\
     A.M.Polyakov, Gauge fields and strings, Harwood Academic,
    1987; \\ A.Yu.Dubin, JETP Lett. 56 (1992) 545;\\
     Yu.S.Kalashnikova, A.V,Nefediev, Phys. At. Nucl. {\bf 60} (1997)
     1389;
     {\bf 61} (1998) 785.

    \bibitem{9}
    R.L.Jaffe, Color, spin, and flavour-dependent forces in quantum
    chromodynamics, hep-ph/0001123, January 2000.

    \bibitem{10}
   B.O.Kerbikov, M.I.Polikarpov and L.V.Shevchenko, Nucl. Phys. {\bf
   B331} (1990) 19;\\ A.M.Badalian, Unified description of ground
   state mesons and baryons in  a potential model, Preprint ITEP-21,
   1987.


  \bibitem{11}
   S.Capstick and N.Isgur, Phys. Rev. {\bf D34}
   (1986) 2809

   \bibitem{12}
   G.Bali, QCD forces and heavy quark bound states, hep-ph/0001312,
     January 2000.

        \bibitem{13}
    Yu.S.Kalashnikova, I.M.Naridetskii, and Yu.A.Simonov, Yad. Fiz.
       {\bf 46} (1987) 1181.

         \bibitem{14} H.Leutwyler, The masses of
     the light quarks, CERN-TH/96-25



     \end{thebibliography}
\end{document}